# Prompted Software Engineering in the Era of AI Models

Dae-Kyoo Kim
Department of Computer Science and Engineering
Oakland University
Rochester, MI 48307

**Abstract**

This paper introduces prompted software engineering (PSE), which integrates prompt engineering to build effective prompts for language-based AI models, to enhance the software development process. PSE enables the use of AI models in software development to produce high-quality software with fewer resources, automating tedious tasks and allowing developers to focus on more innovative aspects. However, effective prompts are necessary to guide software development in generating accurate, relevant, and useful responses, while mitigating risks of misleading outputs. This paper describes how productive prompts should be built throughout the software development cycle.

1. Introduction

Traditionally, software development has been a labor-intensive process that involves a team of developers working together to analyze requirements, design and implement models, test the implementation, and fix errors until the software is ready to be deployed. This process is often time-consuming and expensive.

The emergence of AI models such as ChatGPT [1] has the potential to revolutionize the field of software development. These powerful models can provide significant assistance throughout the software development cycle, using their high-quality domain knowledge learned from vast amounts of data to make predictions and generate software artifacts such as designs and code automatically. This technology can help reduce the time and cost involved in software development, while also improving the quality of the resulting software.

Despite the potential of language-based AI models, their effectiveness heavily relies on the quality of the prompts provided to them. Poorly crafted prompts can lead to negative impact on the software development process, resulting in poor quality software and increased time and cost. The problem, therefore, is to design effective prompts that can guide the AI model to generate accurate, relevant, and useful responses, while mitigating potential risks associated with biased or misleading outputs.

This paper introduces prompted software engineering (PSE), which integrates prompt engineering [2] into software engineering to leverage AI models in software development. Prompt engineering, a technique from the natural language processing field, optimizes input queries to generate more accurate and useful responses from an AI model. Applying this approach to software engineering can enhance software development by enabling the use of AI models to produce high-quality software with fewer resources. AI-generated responses can automate tedious tasks, such as code refactoring, unit testing, debugging, and documentation, enabling developers to concentrate on more sophisticated and innovative software development aspects.

## 2. Prompt Engineering

Prompt engineering has evolved in the area of natural language processing (NLP) to enhance the performance of language models [2-4]. It involves the process of designing and refining prompts or input queries to elicit more accurate, relevant, and useful responses from an AI model. Prompts should be designed with clarity and specificity to be effective. They can be questions, statements, or commands - each serving different purposes. Understanding the context and constraints surrounding the prompt is essential to shaping the AI-generated response. Prompts can be refined through testing to optimize the quality and relevance of AI-generated responses. Refining prompts involves experimenting with different wording, phrasing, and level of detail to determine the most effective prompts, and this process is repeated until the desired output is achieved.

## 3. Prompted Software Engineering

Language-based AI models are popular for their ability to generate high-quality outputs in various applications such as content generation and conversation agents [3]. Effective prompts are crucial to obtaining useful and accurate responses, minimizing misunderstandings, and enhancing user experience. PSE as depicted in Figure 1 leverages prompt engineering and AI models to develop quality software with less cost by generating various types of artifacts and automating routine tasks in software development. This enables developers to focus on more complex and creative aspects of software development, while improving the efficiency of the software development.

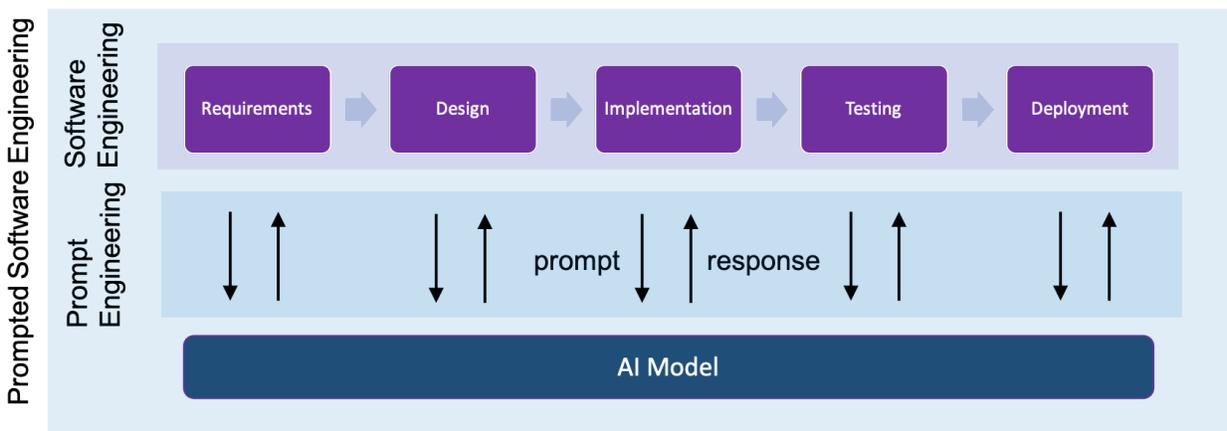

Figure 1: Prompted Software Engineering

In order to devise effective prompts in PSE, following principles should be observed [5].

- *Clarity*: Prompts should be clear and easy to understand. Ambiguous or confusing prompts can lead to misunderstandings and inaccurate or irrelevant responses.
- *Specificity*: Prompts should be specific and provide sufficient details to guide the user through the process.
- *Brevity*: Prompts should be concise and to the point. Lengthy prompts can be overwhelming and confusing.
- *Structure*: Prompts should be well-structured and organized in a logical manner. This can help users understand the process and provide accurate and relevant information.
- *Intent*: Prompts should clearly convey the intended goal or outcome of the process.
- *Context*: Prompts should take into account the context in which they are being used.
- *Constraints*: Prompts should take into account any constraints or limitations that may impact the user's response.

### 3.1 Requirements Phase

The requirement phase is concerned with gathering and analyzing requirements and modeling use cases and domain concepts. Prompts in this phase should take into account the context of the requirements gathering process and any constraints or limitations that may impact the requirements gathering process, while while avoiding technical jargons or complex languages that may be confusing.

**Requirements gathering.** This activity involves identifying stakeholders and conducting interviews, surveys, and workshops to gather their requirements. Prompts can be used in this activity to elicit specific information from stakeholders or users, helping guide the conversation and ensure that all relevant information is collected. Specifically, prompts can be used to:
- Gather requirements by asking about the specific features and functionalities required for the software system. This can help the stakeholders to clearly articulate their needs and ensure that all requirements are captured.
- Explore the context in which the software system will be used. This can involve asking about the user's business objectives, budget, and timeline, which helps to identify any potential issues or limitations that need to be considered during the development process.
- Evaluate the stakeholders' priorities and preferences. This can involve asking about the most critical features or functionalities required for the system.

**Requirements analysis.** The goal of this activity is to identify the essential features, functionalities, and qualities needed to meet the stakeholders' needs. Prompts in this activity can be used to:
- Clarify certain aspects of the requirements, such as the intended users, the expected inputs and outputs, and the specific functionality required.
- Provide clear and concise instructions as to how the necessary information can be provided. This involves breaking down complex requirements into smaller, more manageable parts and providing prompts for each part. For example, a prompt could be designed to guide stakeholders through the process of creating user stories.

**Use case modeling**. This activity focuses on developing use case models to illustrate how the software will be used by end-users (actors) in terms of system operations and interactions between the actor and the system. Prompts in this activity can be used to
- Elicit detailed descriptions of a use case, including the actors involved, the preconditions, the flow of events, and the post-conditions, for example, "What are the preconditions for this use case?"
- Identify the different scenarios that may arise during the use case, for example, "What happens if the user enters incorrect data?"
- Refine the flow of events in a use case, for example, "Can you elaborate this step?" or "Is there another way to achieve this goal?".
- Validate the assumptions made in a use case, for example, "Are there any other actors involved in this use case?".

**Domain modeling**. This activity aims to identify the real-world concepts involved in the system and their relationships based on use case specifications. This involves identifying domain classes and their attributes, and defining the relationships between them. Prompts in this activity can be used to:
- Elicit specific information about domain classes and their attributes such as the scope and purpose of each class and the description of attributes.
- Establish the relationships between domain classes and their types (e.g., association and generalization).
- Specify multiplicities on association ends of domain classes. For example, prompts could ask for assessing the appropriateness of multiplicity, detecting inconsistencies among multiplicities and between multiplicities and requirements, identifying potential multiplicities on data relationships, and evaluating the impact of changing multiplicities. This task can also lead to identifying missing requirements on data relationships or constraints.

**3.2 Design Phase**

The design phase focuses on creating solutions for the requirements gathered in the previous phase. This phase involves designing architectures and components, and making various design decisions.

**Architecture modeling.** This activity focuses on creating a high-level view of the system's structure, including the major components and their relationships based on requirements and domain models, utilizing architectural patterns [6]. Prompts in this activity can be used to:
- Identify the key components based on requirements and the relationships between them, and creating documentation for components.

- Recommend suitable architectural patterns (e.g., layered, pipe-filter, client-server, and publisher-subscriber) for the system and evaluating them based on their advantages and disadvantages, assessing the suitability of a particular architectural pattern, evaluating trade-offs between different patterns in terms of qualities (e.g., maintainability, scalability, and performance), and analyzing the potential impact of adopting a pattern on the system.

**Component modeling.** The aim of this activity is to develop the detailed design of each component in the architecture. This includes specifying the functionality of components in terms of design classes and their interactions by making various design decisions such as such as determining types of attributes, operations, object interactions, and navigability. Prompts in this activity can be used to
- Identify the necessary design classes for individual components and specify the attributes and operations that each class should possess.
- Define the relationships between the classes and their types (e.g., association, aggregation, composition) along with navigability on relationship ends.
- Apply design principles [7] such as high cohesion and low coupling to ensure the quality of components. For example, prompts can ask about the responsibilities of component elements and how they relate to each other, ensuring that the component is focused. They can also ask about the relationships between component elements and suggest ways to minimize those relationships, reducing coupling and improving maintainability.
- Guide the application of design patterns [8] (e.g., singleton, façade, and observer) such as recommending suitable patterns to a particular problem or context, evaluating the benefits and drawbacks of patterns, incorporating patterns, and assessing the impact of patterns after application.

### 3.3 Implementation Phase

The goal of the implementation phase is to implement the design by writing code and evaluating conformance to the design. It also involves refactoring code as necessary to improve its structure and maintainability and documenting the code and its functionality.

**Writing code.** This activity aims to implement the design in a chosen programming language, while observing any required guidelines. Prompts in this activity can be used to:
- Select appropriate data structures and algorithms based on the design and constrains imposed by quality requirements (e.g., efficiency, performance).
- Follow coding standards and guidelines such as naming conventions, formatting, and documentation, assess the readability of the code, and identify any violations of coding standards.
- Conduct code reviews to evaluate the quality of the code, identify any potential issues or bugs, and recommend improvements.

**Code refactoring.** This activity focuses on improving the quality of the code by restructuring it without changing its external behaviors. Prompts in this activity can be designed for providing guidance on which refactoring techniques to apply in specific situations and helping to identify potential issues or risks associated with code changes. Some specific examples of prompts for code refactoring could include:
- Suggesting specific refactoring techniques [9] based on the code being reviewed to improve the code's structure and readability. For example, a prompt could suggest that a method be extracted from a larger block of code and turned into a separate function.
- Highlighting code smells and potential refactoring opportunities. Prompts can be used to identify areas of code that could be improved through refactoring by pointing out common code smells and issues. For example, a prompt could identify duplicated code or overly complex code that might benefit from refactoring.
- Providing guidance on how to use refactoring tools. Refactoring can often be facilitated through the use of specialized tools that automate certain aspects of the process. Prompts can provide guidance on how to use these tools effectively and efficiently.
- Evaluating the impact of code changes. Refactoring code can sometimes have unintended consequences, such as introducing bugs or causing performance issues. Prompts can be used to help evaluate the potential impact of code changes before they are implemented, by simulating the effects of the changes and suggesting potential risks or issues.

**Conformance evaluation.** Ensuring that the implementation conforms to the design specifications. Prompts can be used to support conformance evaluation by checking whether the implementation adheres to the design specifications. Some of the tasks that can be accomplished using prompts include:
- Reviewing the design specifications and compare them with the implementation to identify any deviations.
- Performing a walkthrough of the implementation to verify that it satisfies the design specifications.
- Verifying input/output data, program flow, and error handling against the design specifications.
- Logging and tracking any discrepancies found during conformance evaluation and prompt the developer to resolve them.

**Documentation.** This activity is concerned with creating documentation for the code, such as writing comments in code and documenting functions or class interfaces. Prompts in this activity can be used to:
- Write comments in code. Prompts can be designed to ensure that the comments adequately describe the purpose and functionality of the code. For example, prompts could ask developers to explain the intent of a piece of code, provide background information on the rationale behind the code, or suggest best practices for commenting code.
- Document functions or class interfaces. Prompts can be used to specify the input and output of each function, explain its functionality, and provide examples of how to use it.

## 3.4 Testing Phase

The goal of this phase is to ensure that the software product meets the specified requirements and has minimal defects. The major activities involved in this phase include test planning, test design, test execution, defect tracking and reporting, and test results analysis.

**Test planning**. In this activity, the testing team defines the scope, objectives, and testing approach for the project. The team identifies the types of testing to be conducted, the testing tools to be used, and the resources required for testing. Prompts in this activity can be used to:
- Define the scope of testing in terms of features, functions, and modules and setting priorities based on the criticality of each.
- Establish objectives of testing. The objectives may include identifying defects, assessing the quality of the software, and ensuring that the software meets the requirements.
- Identify testing approach. The team can choose between manual or automated testing, black box or white box testing, and other testing approaches (e.g., gray box testing) depending on the scope and objectives.
- Identify testing types to be conducted. The team can choose between unit testing, integration testing, system testing, and acceptance testing depending on the software development lifecycle.
- Identify testing tools that best fit the testing scope, objectives, approach, and type.
- Identify testing required for testing such the number of testers required, the hardware and software requirements, and the budget for testing.

**Test design.** This activity focuses on creating a set of test cases that cover the requirements of the software. Prompts in this activity can be designed to:
- Ensure the testability of the requirements in terms of completeness, consistency, clarity, and accuracy.
- Design test cases to cover all the functional and non-functional requirements, effectively detect defects, and be efficient in terms of time and resources.
- Ensure that the test data used in the test cases is valid, relevant, and sufficient. Some examples include "Have all the possible scenarios been covered?" and "Are the test data values within the acceptable range?"

**Test execution.** In this activity, the testing team executes the test cases on the software product to identify defects or bugs. Prompts in this activity can be used to:
- Prioritize test cases based on their importance and likelihood of failure (higher the likelihood, the more critical).
- Identify the necessary input data required for executing each test case.
- Identify which test cases need to be re-executed after a defect is fixed (regression testing), ensuring that previously working functionality has not been impacted by the changes made to fix the defect.

- Ensure that the testing team has access to the appropriate test environment for executing the test cases. This includes ensuring that the necessary hardware, software, and network configurations are available.

**Defect tracking and reporting.** This activity is concerned with tracking the status of the defects found during testing and reports them to the project team. The team prioritizes the defects based on their severity and impact on the software product. Prompts in this activity can be used to:
- Capture relevant information about defects, such as severity, steps to reproduce, and the test case in which they were found. This information can be logged into a defect tracking system for the development team to analyze and fix.
- Report defects with a clear and concise description, severity level, and impact on the software product.
- Prioritize defects based on their severity, the likelihood of occurrence, and impact on the software product.

**Test results analysis:** The testing team analyzes the test results to identify patterns and trends in defects. The team uses this analysis to improve the quality of the software product by identifying areas for improvement. Prompts in this activity can be used to:
- Identify commonalities between defects, such as the module or component of the software where most defects are found, or the types of defects that occur most frequently. This information can help the testing team to focus their efforts on areas that are more likely to contain defects.
- Identify the underlying causes of defects. By understanding the root causes of defects, the team can take steps to prevent similar defects from occurring in the future.
- Evaluate the effectiveness of the testing process. For example, the team can use prompts to compare the number of defects found during testing to the number of defects that were expected to be found. This can help the team to identify areas for improvement in the testing process.
- Assess the extent to which the software has been tested. The testing team can use prompts to identify gaps in the testing process and take steps to address them.

### 4.5 Deployment Phase

This phase is the final phase of the software development life cycle (SDLC) and involves the actual release of the software product to end-users. The main objective of this phase is to ensure that the software product is delivered and installed correctly, and that it functions as intended in the production environment. The major activities in this phase include release planning, deployment and installation, configuration and customization, monitoring and maintenance. Prompt in this phase can be used to:

- Identify the stakeholders involved in the release planning phase and their roles and responsibilities. The prompts can also help in creating a comprehensive release plan that includes the release schedule, the deployment process, and the required resources.

- Identify the deployment and installation requirements for the software product. The prompts can help in determining the hardware and software requirements, installation procedures, and any dependencies that need to be resolved.
- Identify the configuration and customization requirements for the software product. The prompts can help in determining the user preferences, software configurations, and any customization needed for specific user groups.
- Identify the monitoring and maintenance requirements for the software product. The prompts can help in setting up monitoring tools and procedures to ensure that the software product is performing as expected, and in identifying any issues that require maintenance or updates.

## 4. Case Example

We conducted a case study using ChatGPT to develop an online tour reservation system (TORS) from the requirement phase to the implementation phase [10]. ChatGPT demonstrated impressive performance throughout the study, successfully identifying and clarifying ambiguities in the TORS requirements, extracting comprehensive requirements, and generating well-structured use case specifications. However, some of the clarifications inferred by ChatGPT involved important business decisions to be made, which requires thorough review. ChatGPT showed its ability to identify domain concepts and attributes, although some inferred attributes could bring potential complexity. In the design phase, ChatGPT was able to make various design decisions (e.g., identifying operations) and create behavioral models with logic. In the implementation phase, ChatGPT generated a basic implementation that included attributes and empty methods and demonstrated the ability to fix errors precisely and efficiently. However, inconsistencies among produced artifacts remain to be improved.

## 5. Conclusion

In conclusion, the growing impact of AI models across various fields has extended to software engineering, presenting opportunities for the development of high-quality software at lower costs. It is imperative that the practice of software engineering adapts to this growing trend to take advantage of the benefits AI offers. This, in turn, will create new fields in software engineering, enabling human developers to focus on more creative activities. However, it is worth noting that PSE is still in its early stages and requires continuous evolution to fully realize its potential.